# An Automated Attendance System based on NFC & X Bee Technologies with a Remote Database

**Ezer Osei Yeboah-Boateng, Emmanuel Owusu Asamoah, Vera Dzidedi Segbedzi**

*Abstract—The main aim of this research was to automate attendance registration, thereby reducing human involvement in the whole process. Typically, the system works by storing vital staff personable information, such as Name, Job specification, etc. into a MySQL database upon engagement. The staff is identified with a unique key associated with an NFC based ID card within the database. So on typical work day, an employee scans his/her ID card on the PN532 reader in close proximity. The exact time and date, together with the unique identifier of the scanned card are stored locally on a storage media, before the data is relayed via the XBee to the remote database. The captured data is then authenticated by comparing with the pre-entered data to give access or authorization to the corporate resources, as well as recorded for attendance purposes. Our experiment shows that the automated attendance system is more effective, efficient and reliable, due to its real time capability, remote monitoring and attendance reports that it provides to the institution.*

## I. INTRODUCTION

Today attendance has become an important part of every organization/institution. Recording and monitoring of staff attendance is an area of administration that requires significant amounts of time and efforts in a working environment. In order to maximize performance in any organization, staff irrespective of their positions are required to give off their optimum output. This output is coupled with high productivity hence the need to monitor the commencement and closure of business activities. Attendance can be defined as the action or state of going regularly to or being present at a place or event[1]. Staff attendance tracking is a common practice in almost all organizations in order to maintain their performance standards. We are aware of the various types of attendance systems that have been developed. Attendance systems using punch cards, log books, fingerprint systems, barcodes, QR codes and also RFID are available but can still cause lots of problems such as providing incorrect information to the users[2][3]. The purpose of the automated attendance system is to computerize the traditional way of recording attendance and provide an efficient and automated method to track attendance in institutions [4].

This paper presents the integration of a computing system at workplaces using NFC and XBee technology. NFC technology is a powerful tool used to solve not only the issues related to attendance. Its peer-to-peer property also allows two powered devices to exchange information when they touch each other[5]. XBee is a brand of radio that supports a variety of communication protocols, including ZigBee, IEEE-802.15.4, and Wi-Fi, among others [6]. In addition a real time intelligent system is implemented in conjunction with NFC hardware to record attendance before, during and after working hours in institutions. NFC technology is a short range radio technology that enables communication between devices that either touch or are momentarily held close together. It provides bidirectional communication over a short distance of about ten centimetres[7]. A visit to a firm/organization, can tell based on observable organizational culture how staff members register their presence or otherwise. For manual and laborious systems, they are usually characterised by unreliable time records, low productivity as well as impact on the bottom-line. Many institutions are saddled with the challenge of effectively monitoring staff time and attendance records. An automated time and attendance system that presents reliable accuracy, increased productivity and bottom-line saving is worth having in the organization.

The proposed system is based on a passive NFC card which interacts with a scanner at the entrance of every institution. The reader generates an electromagnetic field which powers the passive card to enable it to exchange data in a form of a radio signal. The radio signal is interpreted by the scanner which sends the data out to a processing system. The interpreted radio signal (data: serial number, text message etc.) is relayed to a wireless module. The wireless module transmits the processed information to a database for the purpose of storage. The objective of this project is to track attendance of staff during working hours using NFC technology. The data collected would be relayed through a point to multipoint network and also stored in a database for individual processing and for the purpose of storage. The implementation of this project will help maximize performance in Institutions and Industries since the system of checking attendance will be properly automated and monitored hence saving time. Budget allocation will become easy since the company would be properly informed of the exact number of staff that are present in real time and their input at work.

## II. REVIEW OF RELATED WORKS

In [8], a tag scanning module and a tag programming module were used to track school fees arrears. The aim of the project was to reduce the cost that was incurred by the terminal printing of pass cards during exam periods. This aim was achieved as there was a drastic reduction in the printing of pass cards by the university authorities. This was done using the tag scanning module in scanning the RFID tags of students and was programmed with the programming module which was responsible for writing and saving data on the RFID tags. The result from the reader was then displayed on an LCD.

Authors in [9] authenticated the RFID tags by developing a database to store the information of students. The researchers did this by designing an attendance system which sent attendance reports of students to parents by the use of GSM. The aim of this project was to monitor student attendance by using RFID in administration. Their objective at the end of the project was achieved as the designed system involved the use of RFID tags, microcontroller, database and a GSM module. The reader picks up data from the RFID tag when it is in close proximity with it. A microcontroller attached to the reader sends the data obtained from the reader to a PC which was interfaced with a software application and a GSM module.

In this design, the researchers [9] authenticated the RFID tags by developing a database to store the information of students. Also, they employed the use of a GSM module which relayed attendance report towards' parents.

Comparing project [8] and [9], project [9] developed a database to store the results for future references and a GSM module to inform parents about their wards attendance in school.

From the above review, it is clear that the methods applied in [9] were much better than [8] because of the database and GSM features involved.Our approach of automating the attendance system is more effective, efficient and reliable due to its real time ability, remote monitoring and attendance reports that it provides to institutions.

### A. XBee& ZigBee Technologies

XBee is a brand of radio that supports a variety of communication protocols, including ZigBee, IEEE-802.15.4, and Wi-Fi, among others. ZigBee is a standard communications protocol for low-power, wireless mesh networking. It is the name of a standard that specifies the application layer of a wireless personal area network in a small area and a low communication rate. The ZigBee standard operates on the IEEE-802.15.4 physical radio specification and operates in unlicensed bands including 2.4 GHz, 900 MHz and 868 MHz, a near-field small scale network is easily configured using ZigBee modules [6].

### B. XBee Module (Series 2)

To check if the radio links were feasible, we decided to use an XBee SERIES 2 module, a free tool for the design and simulation of wireless systems. The XBee ZB modules extend the range of the network through routing[10]. They form self-establishing, self-healing networks for moving data across the network as shown in Figure 1-8. While these modems use the IEEE-802.15.4 protocol for point-to-point communications, higher protocols are placed on top of these for network routing[11].

While distance specifications like 1 mile sound great, there are many factors that can affect the transmission distance including absorption, reflection and scattering of waves, line-of-sight issues, antenna style and frequency. The sensitivity of the receiver allows it to receive and use signals as low as -100dBm, which is 0.1 Pico watts of power. The RF signal uses Direct-Sequence Spread Spectrum, spreading the signal out over the frequency spectrum greatly improving the signal-to-noise ratio (SNR)[12].

We employ the use of the Series 2 in our project because it reports the RSSI level (Receiver Signal Strength Indication) in several ways so that we can monitor the strength of the signal and the simple passing of data between nodes for microcontrollers. The XBee has features that aid in process monitoring and control.

### C. Near Field Communication (NFC)

NFC is the youngest and cleverest member of the family of identification cards. Contactless as defined by the Smart Card Industries applies to short distance communications between two devices that are not physically connected. This permits the development of Contactless services. NFC is one of the many variations of contactless technology that exist today[13]. NFC which means Near Field Communication is a short-range radio technology that enables communication between devices that either touch or are momentarily held close together. It provides bidirectional communication over a short distance of about ten centimetres. The key to successful introduction of NFC technology is to ensure interoperability between NFC devices to maximize the attractiveness of services, and this requires standardization of NFC technology. The advantages mentioned in this section have prompted the use of NFC to track attendance in institutions.

This section examines the pros and cons of technologies in this area, with the aim of selecting the most appropriate technology that can be used for automating attendance in institutions.

The various technologies are compared above with respect to some of the vital parameters concerned with automating attendance in institutions.

**Table 1 Comparing the Various Wireless Sensors and their Applications**

| Parameters | QR-Code | Biometric | RFID | NFC |
|---|---|---|---|---|
| Maximum Operating Range | 1ft | - | 3 m | 4-10 cm |
| Operating Frequency | Varies | Varies | Varies | 13.56 MHz |
| Directional Communication | One way | One way | One way | Two way |
| Bit Rate (Kbps) | 3 | - | Varies[1] | 106/212/424 |
| Data density(bytes) | High data density | High data density | Very high data density | Very high data density |
| Wear and tear of data carrier | High | Not applicable | No influence | No Influence |
| Purchasing cost | low | high | low | low |
| Operating cost | Low | high | none | none |
| Distant between reader and data carrier in centimetres | 1.5-2.5 m | Direct contact | Depends on frequencies used | 10 cm |
| Influence of dirt | Very High | High | No Influence | No Influence |

### D. How NFC Operates

NFC's are based on RFID (Radio Frequency Identification) technology. NFC operates like the Bluetooth technology but it is more efficient in its operations. This feature allows it to be low-powered, and to avoid interference with other radios built into devices using it. Near field communication (NFC) is a wireless communication technology that operates within the globally available and unlicensed ISM (industrial, scientific, and medical) band of 13.56MHz with data transfer rates of up to 424 Kbit/s [13]. This fast and easy communications between all types of wireless devices through touch and go convenience make NCF's the perfect solution for controlling data in our increasingly complex and connected world.

NFC is an extension of RFID. NFC exchanges also involve an initiator and a target like RFID. However, it can do more than just exchange UIDs and read or write data to the target. The primary difference between RFID and NFC is that, NFC targets are often programmable devices, like mobile phones. [13][7]. This means that rather than just delivering static data from memory, an NFC target could actually generate unique content for each exchange and deliver it back to the initiator[7]. NFC devices have two communication modes, which are the active or passive mode. The passive mode is responsible for achieving significant power savings and extending the battery life while the active mode provides all the power needed for communication with passive devices through their internally powered RF field[14]. This process powers the contactless smart cards which ensure that data remains accessible even if the NFC enabled device is switched off. For NFC communication to be possible, one device has to act as reader and another as the tag. The reader is the active device which generates RF signals to communicate with the tag, while the tag is a thin device which contains an antenna and small memory. RFID operations deal with communication between two active devices.

In addition, while contactless smart cards, only support communication between powered devices and passive tags, NFC's provide peer to peer communication. Thus, NFC's combine the feature to read out and emulate RFID tags and in addition share data between electronic devices that both have active power. NFC's can also be used to initiate WLANs, Bluetooth and other wireless connections without going through configuration menus[15]. NFC devices have three operating modes which are; readers/writers that read data from a target and write to it; emulators, acting like RFID tags when they're in the field of another NFC or RFID device; operate in peer-to-peer mode, in which they exchange data in both directions.

## III. HARDWARE DESIGN

### A. PN532 NFC Breakout board (NFC Reader)

The PN532 is a highly integrated transceiver module for contactless communication at 13.56 MHz based on the 80C51 microcontroller core. The transmission module utilizes an outstanding modulation and demodulation concept completely integrated for different kinds of passive contactless communication methods and protocols at 13.56 MHz and supports six 6 different operating modes as discussed[16]. There are four types of tags defined by the NFC forum, all based on the RFID control protocols described previously. There is a fifth that is compatible, but not strictly part of the NFC specification. Types 1, 2, and 4 are all based on ISO-14443A, and type 3 is based on ISO-18092[7][16].

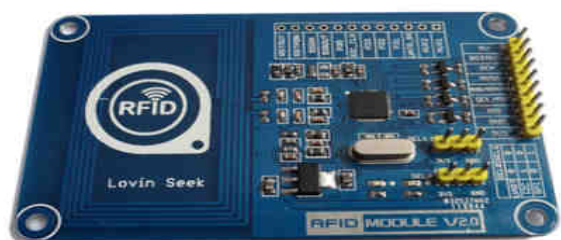

**Fig. 1. Diagram of the PN532 Breakout board Source: [16]**

In this project we chose the Mifare Classic tags (ISO-14443A) because it is readily available, most common in the contactless tag environment and also compatible with most NFC systems[7][16].

### B. MIFARE Classic Cards 1K

MIFARE Classic cards come in 1K and 4K varieties. While several varieties of chips exist, one main chipsets (MIFARE MF1S503x) used in this project is described: The MF1S503x is used in a contactless smart card according to ISO/IEC 14443 Type A developed by NXP semiconductors[17]. It is used in applications like public transport ticketing and other applications. In terms of this project we use the MF1S503x as a contactless smart card bearing the data of the respective card holder. MIFARE Classic cards typically have a 4-byte NUID (Non Unique Identifier) that uniquely (within the numeric limits of the value) identifies the card.

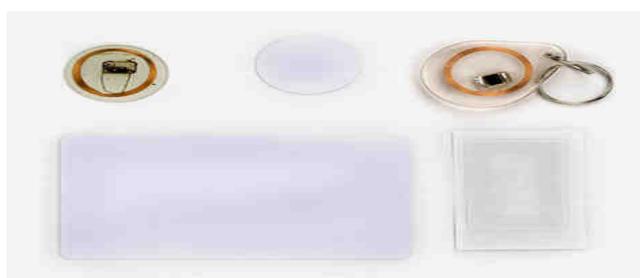

**Fig.2. Different Types of MIFARE Classic Cards Source: [18]**

### C. Arduino Mega

The Arduino Mega is a microcontroller board based on the ATMEGA2560 (datasheet). It has 54 digital input/output pins of which 14 can be use as Pulse Wave Modulation (PWM) outputs. It also includes sixteen (16) analogue inputs, four (4) UARTs (hardware serial ports) and also a 16 MHz crystal oscillator, a USB connection, a power jack, an ICSP header, and a reset button [19][20]. In the design of this project we chose the Arduino Mega because of its enormous pin usage and flexibility. It is easy to use and also inexpensive with respect to the demands of the project.

### D. SD Shield 3.0

SD Shield 3.0 is a break out board for standard SD and TF card. It is compatible with 5V and 3.3V Arduino mainboards. The communication between the microcontroller and the SD card uses SPI (Serial Peripheral Interface), which takes place on digital pins 50, 51, and 52 (Arduino Mega). Additionally, another pin must be used to select the SD card. This can be the hardware pin 53 on the Arduino Mega. This module also has a Micro SD socket on the back side. These cards are formatted FAT32 and SD. The SD library in the Arduino development module supports the FAT16 and FAT32 File Systems [21].

### E. DS1307 Serial Real Time Clock

DS1307 serial real-time clock (RTC) is a low power, full binary-coded decimal (BCD) clock/calendar plus 56 bytes of NV SRAM. In addition, address and data are transferred serially through an I2C, bidirectional bus. Also, the clock/calendar provides seconds, minutes, hours, day, date, month, and year information. The end of the month date is automatically adjusted for months with fewer than 31 days, including corrections for leap year. The clock operates in either the 24-hour or 12- hour format with AM/PM indicator [22]. The DS1307 has a built-in power-sense circuit that detects power failures and automatically switches to the backup supply [22][23]

### F. Arduino Ethernet Shield

Arduino Ethernet Shield allows an Arduino board to connect to the internet. It is based on the Wiznet W5100 Ethernet chip (datasheet). The Wiznet W5100 provides a network (IP) stack capable of both TCP and UDP. It supports up to four simultaneous socket connections. The Ethernet shield connects to an Arduino board using long wire-wrap headers which extend through the shield. This keeps the pin layout intact and allows another shield to be stacked on top [24]. The Ethernet Shield has a standard RJ-45 connection, with an integrated line transformer and Power over Ethernet enabled. There is an on-board micro-SD card slot, which can be used to store files for serving over the network [24].

### G. Conceptual Design

A systematic approach is used to bring together all the components discussed in the previous sections to automate the attendance system using NFC Technology. A Complete block diagram is provided below to show the entire flow of the attendance system. This will give a clear diagrammatic representation of how the various parts of the system functions.

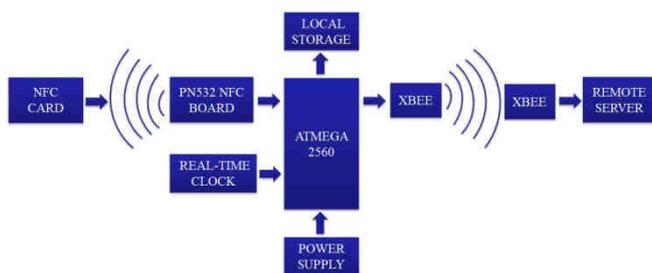

*Source: Field Work*

**Fig. 3. Conceptual Schematic of NFC and XBee based Attendance System**

The entire system is triggered when the NFC Card is brought into close proximity (5cm) with the NFC reader or scanner (PN532). The reader generates an electromagnetic field which causes electrons to move through the tag's antenna and subsequently power the chip embedded in the NFC Card. The Card which is a passive card (has no power source of its own) is powered wirelessly by radio waves emitted from the reader/scanner (PN532). The powered chip inside the NFC Card then responds by sending its stored data back to the reader in the form of another radio signal.

The radio signal is interpreted by the reader which sends the data out to the ATMEGA 2560 (Arduino Mega) microcontroller. The interpreted radio signal (data: serial number, identification number) together with the real time generated by the real time clock at the exact time the tag is scanned is relayed serially to the local storage (SD Card) as a backup and also to the XBee module. The XBee module being a radio module transmits the processed information wirelessly to a remote receiving XBee. The information received by the receiving XBee is relayed serially to a computer (Remote Server) for the purpose of storage. All these components work together with respect to the design to achieve the transmitting end of the automated attendance system. The transmitting end basically reads card details with key attention to the time and date the card was read as these details are primarily saved in a local storage system for later assessment. These details are then relayed wirelessly to a remote server for remote storage.

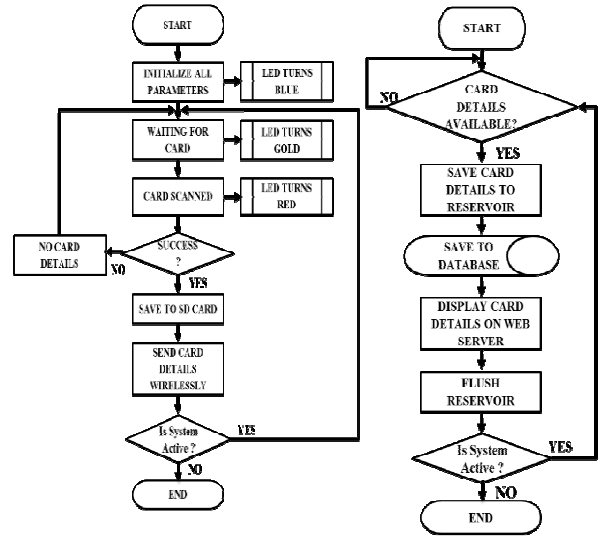

**Fig. 4. Flow Chart of Transmitting and Receiving End Source: Field Work**

## IV. RESULTS AND ANALYSIS

As per the objectives this project sought to achieve, tests were conducted at various stages of the project work and the results of these tests are detailed in this section. At the beginning of chapter 3, the project was split into three sections: the hardware, the wireless and the software sections. Tests were carried out at the end of each section before proceeding to the next stage and the tests we run sought to determine the following:

1. The speed with which one can check into an office with our NFC tag.
2. The speed with which data on the card can be uploaded and relayed to a remote server.
3. The error rate of the reader.
4. The error rate of the remote server.

### A. Test Results of PN532 Reader

The tests were carried out on a sample space of seven NFC tags. First and foremost the PN532 reader is placed at the security post of the office, where the attendance is recorded. Next, a worker who walks into the premises and scans their NFC cards at the security post as they enter. The scanning of NFC cards is an instantaneous process and the range between the PN532 reader and the NFC card is 5cm.

We sought to study the check in process at offices and schools and thus determine how much time it will take for our PN532 reader to accurately read the results of each card and grant or deny access to the tag bearer as compared to the traditional method. We also sought to determine the accuracy of the PN532 when it was used to scan multiple cards in quick succession and how fast the transmitting XBEE relayed the data to the remote server.

We were limited in our ability to test the accuracy of the PN532 reader when it is used to scan multiple cards in quick succession because of the limited number of tags available for us to use as a sample space. However, we were able to come to some good conclusions with what we had available. The traditional method of appending a signature to your name in a log book took a worker approximately 30 seconds to check into the office premises.

However, using the PN532 reader took exactly two (2) seconds to scan a tag and display its results. This meant that the process had been sped up by more than 93% of the previous time.

**Fig. 5. Test Results of PN532 Display on Serial Monitor Source: Field Work**

Fig. 5.shows the output when a tag is brought into close proximity with the PN532. It displays the card properties, which includes the (UID) User Identity (unique string of characters) as well as the date and time the tag was scanned, and also saves the data to a text file and sends it over to the XBEE. The prints of zero (0) means the system is ready to scan another tag. It takes half a second (0.5 seconds) for the system to scan another tag.

**Table II. Test Results of the PN532 NFC Transceiver Source: Field Work**

| Method | 1 worker | 10 workers | 60 workers | 100workers |
|---|---|---|---|---|
| **Manual Entry (Field Work)** | 30secs | 3mins | 30mins | 50mins |
| **RFID tag (Mensah et al)** | 2secs | 20secs | 2mins | 3mins 30secs |
| **NFC tag (Field Work)** | 0.5secs | 5secs | 30secs | 50secs |

### B. Test Results of the XBee Module

The X-Bee module was tested for its range and speed of data transfer with respect to this project. It was configured with the necessary parameters as discussed using XCTU. The range of testing was done using two X-Bee modules which were mounted on Arduino boards. One X-Bee module was configured as a Coordinator and connected to an Arduino Mega and the other XBee was configured as a Router and connected to an Arduino Uno at the remote end. The range and speed test was accomplished by implementing a chat between the two XBee's and sending a character of strings in the form of text using the API mode (Application Programming Interface) in the XCTU.

We further carried another test to determine the amount of time the XBEE module takes to relay the data received, to a remote server via the router XBEE out on our sample space of 7 tags. The amount of time needed to transmit the information on the tag to the server was 1 second. This amount of time is relatively short and would subsequently prevent queues at the server even if the number of people present to check in at a particular time is a lot.

**Fig. 6. Test Results of XBEE Router "Display on Serial Monitor" Source: Field Work**

Fig. 6.above shows the output of the receiving end of our system. It prints out the tag details it receives from the transmitting end. The tag/card details include the UID and the date and time the tag was scanned. This means the router XBEE is always listening to the coordinator to receive data.

### C. Test Results of the Error Rate of the PN532 Reader

The error rate of the PN532 reader relates to the number of times it gives a wrong reading or makes a mistake when it is used to scan multiple tags in quick succession. Once again we were limited by the number of tags available for our sample space, but we overcame this challenge by using the same tags over and over again in a cycle in quick succession.

The seven (7) tags were scanned sixty (60) times with an average time interval of 0.5 seconds. After 60 scans in quick succession, the PN532 used an average of 30 seconds to scan the tags which reduced the time of the manual system by 98.33%.

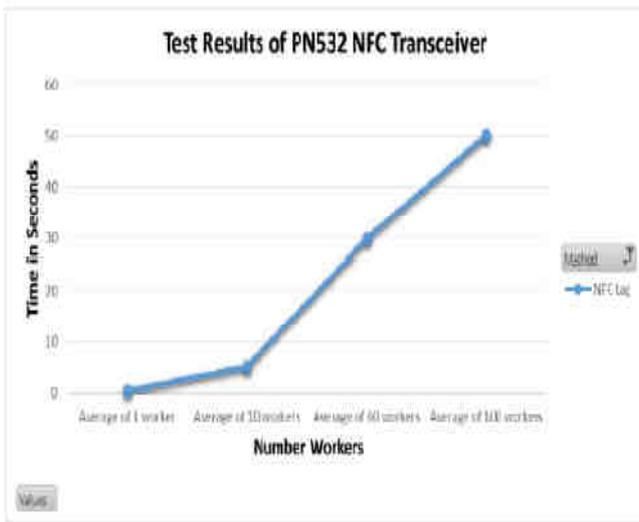

**Fig.7.** **Test Results of PN 532NFC Transceiver Source: Field Work**

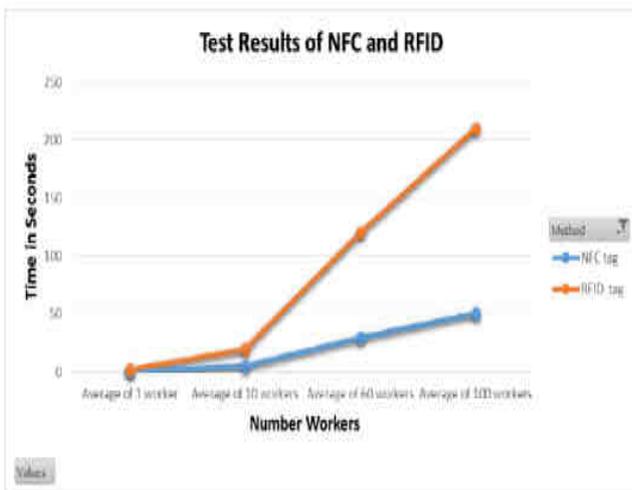

**Fig.8.** **Test Results of NFC and RFID    Source: Field Work**

From Fig. 8.it can be seen that NFC takes a shorter time to scan the tags of 100 workers as compared to RFID.

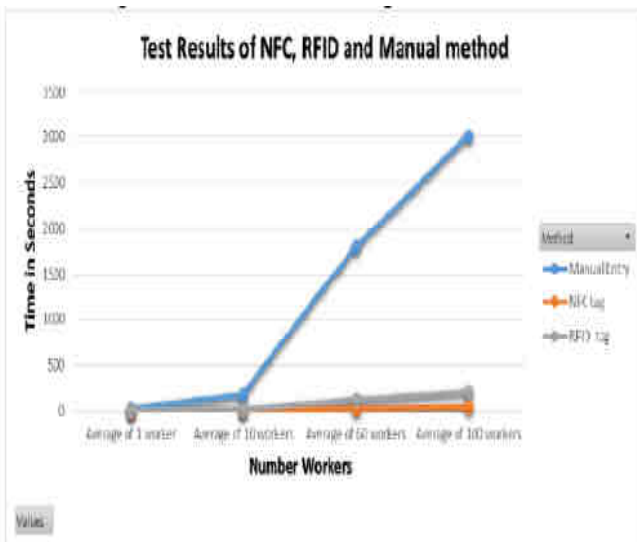

**Fig. 9. Test Results of NFC, RFID and Manual Method Source: Field Work**

From Fig 9.the manual entry mode used a longer time to record the attendance of 100 workers as compared to RFID and NFC.

The above line graphs (Fig. 7, 8, 9) compares the average time taken to scan tags per the number of workers ranging from 1 worker to 100 workers. Three entry methods were mainly used: manual entry, entry using RFID and entry using NFC. Overall, it can be seen that NFC entry used a shorter time period to scan tags of 100 workers as compared the other methods of entry.

## V. CONCLUSION

This project was aimed at automating the attendance system in institutions. We set out to design and construct a system which will minimize the inefficiencies associated with the current system of monitoring the attendance system in institutions with the use of passive electronic cards or tags.

The PN532 transceiver system used for automating the attendance register in this project can scan a single NFC tag at a time.  This is due to the read range of the reader being about 5cm. Once the NFC tag is scanned at this read range by the PN532 reader, the data from the NFC tag gets transferred to the local storage and also sent wirelessly to the remote server. After an in depth study into the various RFID readers, the PN532 reader module was selected as the best for this project. The module was consequently programmed to serve as an interface between the reader and the tag. Configuring the PN532, the Real Time Clock, SD Shield and the XBee (Coordinator) to communicate with the host microcontroller (Arduino Uno) posed the first challenge as the digital pins were limited to all the components mentioned earlier and the power supply of 5volts was not enough to drive all the components to work in their full functions and operation. However, after many trials the Arduino Uno was changed into an Arduino Mega which was able to accommodate all the other hardware components because it had more digital pins and several power pins.

The software section of the project was then designed using programming in C++ and MySQL together with an Ethernet Shield. We chose C++ as the most suitable programming language because it was easy to manipulate and was more accommodative. The simplicity and database size in terms of the entire software size of SQL resulted in MYSQL being selected over SQL.

Furthermore, programming offered a second set of challenges for the project as it required knowledge of programming of some of the components involved in this project. We also faced the difficulty of fusing the programming of the real time clock into our codes. This was eventually achieved with the help of knowledge we acquired from our lessons in basic C++ programming. The final and most challenging part of our work was setting up a wireless link between the reader and the server. This was because XBee technology was new and not really patronized locally. Relaying the data received from the reader was difficult initially but we employed the use of XCTU, a free software provided by Digi which was used to configure and manage XBee's and thus used to test our XBee network.  We also watched videos on YouTube involving people who had worked with the device to gain more knowledge on its operations.

However, the final system we came up with is not perfect and has some limitations which be improved upon to provide a more convenient system. Some of these limitations include;

1. The XBee Series 2 operation length was a short range.
2. Our network was a point to point network as a result we were unable explore much of the functionalities of the XBee in the design of Wireless Sensor Networks.
3. The MYSQL connectivity to the Ethernet Shield showed lots of complex IP and Remote settings of which became a difficulty in configuring.

Finally, after months of planning and detailed study of all the components used in this project, the project started functioning exactly as outlined in the project objective.

**Dr. Ezer Osei Yeboah-Boateng, FHEA,** is a senior lecturer and the acting Head (Dean), Faculty of Informatics, at the Ghana Technology University College (GTUC), in Accra. Ezer is a Telecoms Engineer and an ICT Specialist; an executive with over 20 years of corporate experience and about 8 years in academia. He has over 10 peer-reviewed international journal papers to his credit. His research focuses on cyber-security vulnerabilities, digital forensics, cyber-crime, cloud computing and fuzzy systems.

**Emmanuel Owusu Asamoah** is a Telecoms Engineer and a Teaching Assistant at the Faculty of Engineering, Ghana Technology University College (GTUC) in Accra.

**Vera DzidediSegbedzi** is a Telecoms Engineer and graduated from the Ghana Technology University College (GTUC) in Accra.